\documentclass[twocolumn,prl]{revtex4-1}
\usepackage[T1]{fontenc}
\usepackage[latin9]{inputenc}
\usepackage{xcolor}
\usepackage{amsmath}
\usepackage{amssymb}
\usepackage{graphicx}
\PassOptionsToPackage{normalem}{ulem}
\usepackage{ulem}

\makeatletter

\providecolor{lyxadded}{rgb}{0,0,1}
\providecolor{lyxdeleted}{rgb}{1,0,0}

\DeclareRobustCommand{\lyxsout}[1]{\ifx\\#1\else\sout{#1}\fi}

\usepackage{bm}

\makeatother

\begin{document}
\title{Dilute magnetism in graphene}
\author{Frederico J. Sousa,$^{1}$ B. Amorim,$^{2}$ Eduardo V. Castro$^{2,3,4}$}
\affiliation{$^{1}$University of York, Department of Physics, YO10 5DD, York,
United Kingdom }
\affiliation{$^{2}$CeFEMA, Instituto Superior T\'{e}cnico, Universidade de Lisboa,
Avenida Rovisco Pais, 1049-001 Lisboa, Portugal}
\affiliation{$^{3}$Centro de Física das Universidades do Minho e Porto, Departamento
de Física e Astronomia, Faculdade de Ciências, Universidade do Porto,
4169-007 Porto, Portugal}
\affiliation{$^{4}$Beijing Computational Science Research Center, Beijing 100084,
China}
\begin{abstract}
The phase diagram of graphene decorated with magnetic adatoms distributed
either on a single sublattice, or evenly over the two sublattices,
is computed for adatom concentrations as low as $\sim1\%$. Within
the framework of the $s$-$d$ interaction, we take into account disorder
effects due to the random positioning of the adatoms and/or to the
thermal fluctuations in the direction of magnetic moments. Despite
the presence of disorder, the magnetic phases are shown to be stable
down to the lowest concentration accessed here. This result agrees
with several experimental observations where adatom decorated graphene
has been shown to have a magnetic response. In particular, the present
theory provides a qualitative understanding for the results of Hwang
\emph{et al.} {[}Sci. Rep. \textbf{6}, 21460 (2016){]}, where a ferromagnetic
phase has been found below $\sim30\,\text{K}$ for graphene decorated
with S-atoms.
\end{abstract}
\maketitle


\section{Introduction}

Graphene is the host of many unconventional properties \citep{NGPrmp,katsnelsonBook2012},
the most well known, arguably, being the ultrarelativistic behavior
of the charge carriers around the neutrality point \citep{KNssc07}.
The observation of Klein tunneling physics \citep{YK09} and of a
room temperature quantum Hall effect \citep{NJZ+07}, direct consequences
of the low energy behavior, are examples of a dichotomous interest
which appeals equally to fundamental physics and technological exploitation.
Intrinsic magnetism is, however, a notable missing item on the list
of relevant properties. Two-dimensional magnetic materials have been
discovered recently \citep{zhangLoui2Dmag2017,xiaoJarilloCrI3Mag2D},
but the quest for metal-free, carbon based structures exhibiting long
range magnetic order \citep{MPbook06} is still open, and graphene
is still a possible choice \citep{Tucek2017}.

Magnetic moments have been observed in graphene, either associated
with structural defects \citep{ESH+03,OTH+07,roomTfer2009,NST+12,Birkner2013},
or with the presence of certain adatoms, such as H \citep{xie2011room,McCreary2012,giesbers2013interface,gonzalez2016atomic},
F \citep{Hong2012,NST+12}, S \citep{lanzara2016}, Au \citep{OFarrell2016},
or even certain molecules \citep{Tucek2017,miranda13}. The theory
describing magnetic moment formation in graphene and/or the coupling
to magnetic adatoms has also been studied at length \citep{Duplock2004,Lehtinen2004,Yazyev2007,Boukhvalov2008,Palacios2008,Stauber2008,UYT+09,Castro2009,Yazyev2010,Ding2011,Wehling2011,Castro2011a,McCreary2012,Santos2012,Sofo2012,Rudenko2012,Eelbo2013,Garcia-Martinez2017,Agarwal2017}.
Although much harder to establish than the presence of magnetic moments,
the observation of magnetic long range order has been reported in
several experiments related to the presence of adatoms or molecules
\citep{xie2011room,miranda13,giesbers2013interface,lanzara2016,Tucek2017}. 

The underpinning mechanism leading to long range ordering of a small
fraction of magnetic impurities (or adatoms) in a (semi)conducting
matrix is known: the interplay, via an $s\text{-}d$-type interaction,
between the itinerant electrons and the local magnetic moments effectively
couples the impurities giving rise to an RKKY-like interaction \citep{Ruderman1954,Kasuya1956,Yosida1957};
this interaction between magnetic moments is responsible for the long
range magnetic order below some critical temperature. This mechanism
is behind the magnetic behavior of \emph{dilute magnetic semiconductors}
\citep{Dietl2014}, both in two-dimensions \citep{dasSarma2dDMS}
and in three-dimensions \citep{Konig2000,Berciu2001}. In graphene,
the same principle should be at play, even though other ingredients
may favor or disfavor the magnetic ordering. For example, the ordering
of magnetic moments in hydrogenated graphene \citep{xie2011room}
is known to have a positive contribution from the carbon buffer layer
in graphene grown on SiC \citep{giesbers2013interface}.

The RKKY interaction in graphene has been studied extensively \citep{brey2007diluted,Saremi2007,Black-Schaffer2010,Sherafati2011,Kogan2011,Sherafati2011a,Power2013,Kogan2013,Agarwal2017}.
It couples magnetic moments on the same sublattice ferromagnetically,
and antiferromagnetically on different sublattices. In two-dimensions,
particularly in graphene, the ordering of a diluted set of magnetic
atoms coupled via RKKY interaction has been shown to be possible \citep{Rappoport2009,fabritius2010,Szalowski2014,Moaied2014}.
However, the use of an effective RKKY-type interaction, often derived
perturbatively, carries the disadvantage of losing information about
the underlying electronic system. In fact, both the random positioning
of the adatoms, and the thermal fluctuations in the direction of magnetic
moments, work as sources of disorder to the electronic system. This
impacts the coupling between the magnetic moments, calling for a self-consistent
approach. Such an approach proved essential, for instance, for the
correct understanding of the ferromagnetic properties in diluted magnetic
semiconductors \citep{Berciu2001}, and to study the ordering of magnetic
adatoms on the surface of topological insulators \citep{Abanin2011,Wray2011,Ochoa2015}.

In the present paper, we compute the magnetic critical temperature
for adatoms randomly distributed on a single sublattice, for which
case ferromagnetism is expected, and for a random distribution of
adatoms on both sublattices, which favors an antiferromagnetic behavior.
Coupling mobile electrons and diluted magnetic moments with an $s\text{-}d$
type interaction, we follow a self-consistent approach which takes
into account effects of disorder and is capable of accessing impurity
concentrations as low as $\sim1\%$. We further analyze the impact
of disorder on the electronic sector by inspecting the electronic
density of states and its dependence on relevant quantities such as
adatom concentration, temperature and coupling strength. We further
test the applicability of the present theory by applying it to the
experimental results of Ref.~\citep{lanzara2016}. This work builds
upon and extends previous results where the same type of interaction
has been studied, and effects of disorder partially taken into account
\citep{daghofer-ssibmiig2010,rappoport2011magnetic}. In Ref.~\citep{daghofer-ssibmiig2010}
it was shown that when adatoms are allowed to sit on both sublattices,
the system develops a gap, whereas when only one sublattice is occupied
the system remains gapless, but the spin degeneracy is lifted, allowing
for spin polarization. The magnetic transition has not been accessed,
though. For adatoms in both sublattices, the electronic system develops
a temperature dependent gap and, as shown in Ref.~\citep{rappoport2011magnetic},
adding an external magnetic field leads to a controllable electronic
magnetization. However, in Ref.~\citep{rappoport2011magnetic} only
concentrations above $\sim20\%$ have been considered.

The paper is organized as follows: in Sec.~\ref{sec:Methodology}
we present the methodology followed; the results for magnetic moments
distributed over only one sublattice, are given in Sec.~\ref{sec:single};
in Sec.~\ref{sec:both} we show the results for magnetic moments
distributed over the two sublattices; in Sec.~\ref{sec:Comparison-with-experiment}
we compare our theory with experiments; and conclusions are given
in Sec.~\ref{sec:Conclusions}.


\section{Methodology\label{sec:Methodology}}

\subsection{Model}

The studied system is a graphene lattice with $N=2d^{2}$ carbon atoms
covered with a number $N_{imp}$ of magnetic adatoms, so the impurity
concentration is $x=N_{imp}/N$. We assume that the adsorption occurs
directly on top of the carbon atoms, the preferred position for certain
adatoms \citep{Ding2011}. Electrons in graphene are modeled with
the usual tight-biding Hamiltonian for the honeycomb lattice

\begin{equation}
\mathcal{H}_{\mathrm{TB}}=-t\sum_{\mathbf{R},\boldsymbol{\delta},\sigma}(a_{\mathbb{\mathbf{R}},\sigma}^{\dagger}b_{\mathbf{R}+\boldsymbol{\delta},\sigma}+b_{\mathbf{R}+\boldsymbol{\delta},\sigma}^{\dagger}a_{\mathbb{\mathbf{R}},\sigma}),\label{eq:TB_hamiltonian}
\end{equation}
where $t\approx3\,\text{eV}$ is the hopping coefficient, $\sigma$
is the spin label, and $\boldsymbol{\delta}$ are the displacement
vectors that connect nearest neighbors \citep{NGPrmp}. The operators
$a_{\mathbf{R}}$ ($a_{\mathbf{R}}^{\dagger}$) and $b_{\mathbf{R}}$
($b_{\mathbf{R}}^{\dagger}$) are the electronic annihilation (creation)
operators for the two different sublattices (A and B) acting on the
unit cell with position label $\mathbf{R}$. The interaction between
magnetic adatoms and the itinerant electrons is described by a phenomenological
$s$-$d$ type interaction,

\begin{eqnarray}
\mathcal{H}_{\mathrm{sd}} & = & -\sum_{i=1}^{N_{imp}}\left(J_{\parallel}\mathbf{S}_{i}^{\parallel}\cdot\mathbf{s}^{\parallel}(\mathbf{R}_{i})+J_{z}S_{i}^{z}s^{z}(\mathbf{R}_{i})\right).\label{eq:sd_Hamiltonian}
\end{eqnarray}
This is a spin-spin interaction between the impurities spins $\mathbf{S}_{i}$
and the electrons spins $\mathbf{s}$ on site $i$. We allow for a
spin-anisotropic coupling, which we parameterize via $J_{\parallel}$
and $J_{z}$, respectively, the in-plane and out-of-plane exchange
couplings. We represent the in-plane components of a vector by $\mathbf{v}^{\parallel}=\left(v^{x},v^{y}\right)$.
The adatoms are always assumed to be randomly distributed, either
evenly on both sublattices or only on one sublattice. This allows
us to study effects of disorder in the worse scenario hypothesis,
even though a more realistic model would allow for attraction or repulsion
between adatoms \citep{Shytov2009}. Adding Eqs.~\ref{eq:TB_hamiltonian}
and~\ref{eq:sd_Hamiltonian} together yields our total Hamiltonian:
$\mathcal{H}=\mathcal{H}_{\mathrm{TB}}+\mathcal{H}_{\mathrm{sd}}$. 

We further assume that the adatoms spins are classical, which is justified
within the phenomenological approach taken here (and is always a good
approximation if the spin-$S$ of the impurity is big enough). Doing
this, the problem becomes a single particle problem for the graphene
electrons under the effect of a (disordered) potential created by
the impurities. For each spin configuration of the classical spins,
we can solve exactly the electronic system. So we can integrate out
the electrons and derive an effective Hamiltonian for the classical
spins, $\mathcal{H}_{eff}$, which will be treated within mean field
(MF) theory.

We start by writing the grand canonical partition function $\mathcal{Z}$
as

\begin{equation}
\mathcal{Z}=\int d[S]\,\textrm{Tr}^{(\textrm{Fock})}\,e^{-\beta(\mathcal{H}_{\mathrm{TB}}+\mathcal{H}_{\mathrm{sd}}-\mu\hat{N})}\label{eq: grand canonical partition}
\end{equation}
where $\mu$ is the chemical potential, $\hat{N}$ is the total number
operator, and $\beta=(k_{B}T)^{-1}$. The integral is calculated over
all directions of all impurities spins, $d[S]=d\Omega_{1}d\Omega_{2}\cdots d\Omega_{N_{imp}}$.
Tracing out the electronic states, we obtain

\begin{equation}
\mathcal{Z}=\int d[S]\prod_{n}(1+e^{-\beta(E_{n}(\mathbf{S})-\mu)})=\int d[S]\,e^{-\beta\mathcal{H}_{eff}(\mathbf{S})}\label{eq: grand partition function and effective hamiltonian}
\end{equation}
where we have defined the effective, classical Hamiltonian

\begin{equation}
\mathcal{H}_{eff}(\mathbf{S})=-k_{B}T\sum_{n}\ln\left(1+e^{-\beta(E_{n}(\mathbf{S})-\mu)}\right),\label{eq: effective hamiltonian}
\end{equation}
which is nothing more than the electronic free energy for a given
impurity spin configuration. The energy levels $E_{n}(\mathbf{S})$,
that are the eigenvalues of the total Hamiltonian $\mathcal{H}$,
depend on the spin configuration of the adatoms and are to be obtained
using numerical methods. Equation~(\ref{eq: effective hamiltonian})
can be written in terms of the density of states (DOS) of the system,
\begin{equation}
\rho_{_{\mathbf{S}}}(E)=\sum_{n}\delta\left(E-E_{n}(\mathbf{S})\right),
\end{equation}
as

\begin{equation}
\mathcal{H}_{eff}(\mathbf{S})=-k_{B}T\int dE\rho_{_{\mathbf{S}}}(E)\ln\left(1+e^{-\beta(E-\mu)}\right),\label{eq:effective_hamiltonian_DOS}
\end{equation}
where the information about the classical spins configuration is now
contained in $\rho_{_{\mathbf{S}}}(E)$. 

In the following sections we show the results obtained for this model
in terms of critical temperature ($T_{C}$), obtained through a variational
analysis to be discussed below, and of the DOS. To obtain $T_{C}$
we compute $E_{n}(\mathbf{S})$ using exact diagonalization to solve
the electronic part of the problem in a $d=32$ lattice and performing
1000 disorder realizations. The DOS is calculated by means of the
Haydock recursive method \citep{Haydock2006} for a system a size
of $d=1000$ and 100 disorder realizations. Computing $T_{C}$ using
exact diagonalization (through Eq.~(\ref{eq: effective hamiltonian})
allows us to get much more accurate results compared to using the
DOS (Eq.~(\ref{eq:effective_hamiltonian_DOS}) at the expense of
having to reduce the size of the system.

\subsection{Mean field treatment}

The MF formulation chosen is based on Bogoliubov's inequality \citep{Parisi_book,Chaikin_book,Alonso_2001}
for the system's free energy, $\mathcal{F}$, 
\begin{equation}
\mathcal{F}\leq\mathcal{F}_{MF}+\langle\mathcal{H}-\mathcal{H}_{MF}\rangle_{MF},\label{eq:bogoliubov_ineq}
\end{equation}
where $\mathcal{H}_{MF}$ is the MF Hamiltonian used to calculate
the average $\langle\dots\rangle_{MF}$, and $\mathcal{F}_{MF}$ is
the MF free energy. The tendency for broken symmetry phases can be
studied by minimizing the right hand side of Eq.~\eqref{eq:bogoliubov_ineq}
with respect to the variational parameters of a conveniently chosen
$\mathcal{H}_{MF}$. Here we use the MF Hamiltonian 

\begin{equation}
\mathcal{H}_{MF}=-\sum_{i=1}\mathbf{h}_{i}\cdot\mathbf{S}_{i},\label{eq:MF_Hamiltonian}
\end{equation}
with variational parameters $\mathbf{h}_{i}$ that represent the local
average magnetic field acting on an impurity site $i$. Since we are
interested in magnetic phases where the direction of the impurities
spins depend, at most, on the sublattice they are in, we consider
only cases where $\mathbf{h}_{i}=\mathbf{h}_{A},\mathbf{h}_{B}$ is
constant in each sublattice. 

To give an example, consider the ferromagnetic phase, where all the
spin are expected to point in the same direction. In this case the
simplest approach is to choose $\mathbf{h}_{A}=\mathbf{h}_{B}\equiv\mathbf{h}=h\vec{e}_{z}$,
so that
\begin{equation}
\mathcal{H}_{MF}=-hS\sum_{i=1}^{N_{imp}}\cos\theta_{i}\,,\label{eq:HmfFerro}
\end{equation}
where $\theta_{i}$ is the angle between the impurity spin and $\mathbf{h}$.
The mean field theory now has a single variational parameter, $h$,
whose meaning is clear, even though it may be hard to know the range
of values it can take. However, we can relate $h$ to the magnetization
$m$, defined as

\begin{equation}
m=\frac{1}{N_{imp}}\langle\sum_{i=1}^{N_{imp}}\cos\theta_{i}\rangle_{MF}=\coth y-\frac{1}{y},\label{eq: MF equation}
\end{equation}
with $y=\beta hS$. When all the impurities spins are aligned, $m=1$,
while in the paramagnetic phase, $m=h=0$, giving us two limits for
this order parameter. Note that the averages are easily done in this
formulation since the distribution probability for the orientation
of the decoupled classical spins is known, $e^{-\beta\mathcal{H}_{MF}}/\mathcal{Z}_{MF}$,
where $\mathcal{Z}_{MF}$ is the partition function for $\mathcal{H}_{MF}$.
For a fixed temperature and impurity concentration, we calculate the
free energy using Eq.~(\ref{eq:bogoliubov_ineq}) for different values
of $m$. We obtain a list free energy points that we fit to the polynomial
\begin{equation}
\mathcal{F}=a(T)m^{2}+b(T)m^{4}+c(T)m^{6}.\label{eq: Landau Polynomial}
\end{equation}
Finally, we do the same for different temperatures, which allows us
to determine a critical temperature $T_{C}$ using the relation $a(T_{C})=0$,
characteristic of a second order transition.

\subsection{Long range magnetic order in two-dimensions}

An ordered magnetic phase in our MF approach breaks the continuous
rotational symmetry and, in light of the Mermin-Wagner theorem \citep{Mermin1966},
is ruled out at finite temperatures. We must stress, however, that
due to the 2D nature of graphene, the $s$-$d$ interaction coupling
is expected to be anisotropic on physical grounds. One possible source
for the anisotropy is the fact that, apart from the $s$-$d$ interaction
induced by the magnetic character of the adatom, it should also lead
to spin-orbit like terms which break the SU(2) rotation symmetry of
the electron spin \citep{AiresScatt14,Lado2017}. This has been recently
demonstrated experimentally~\citep{OFarrell2016}. In this case the
theorem does not apply, and it has been shown that 2D long range order
at finite temperatures is stabilized even for a small amount of anisotropy
\citep{dasSarma2dDMS}. 

Moreover, taking as an example the Heisenberg model with long range
interaction decaying as $1/r^{\alpha}$, the Mermin-Wagner theorem
proves the absence of long range order at finite $T$ only if $\alpha>D+2$
\citep{fabritius2010}, $D$ being the dimensionality of the system.
Even though we cannot generally demonstrate that the effective interaction
between magnetic impurities is long range, in the limit where RKKY
model applies we know that the interaction should decay as $1/r^{3}$
\citep{Saremi2007,brey2007diluted,Sherafati2011a,Kogan2011,Sherafati2011,Kogan2013},
which is below the critical $\alpha=4$ in 2D. Even though stronger
conditions exist for oscillatory interactions \citep{Bruno2001},
we note that, for the case of graphene, these oscillations do not
lead to a change of sign of the coupling. Finally, we note that the
spatial decay of the effective magnetic interaction also depends on
the local potential induced by the impurities, as shown recently in
Ref.~\citep{Agarwal2017}. In certain cases, the coupling between
the impurities becomes even more long-raged.


\begin{figure*}
\begin{centering}
\includegraphics[width=0.9\columnwidth]{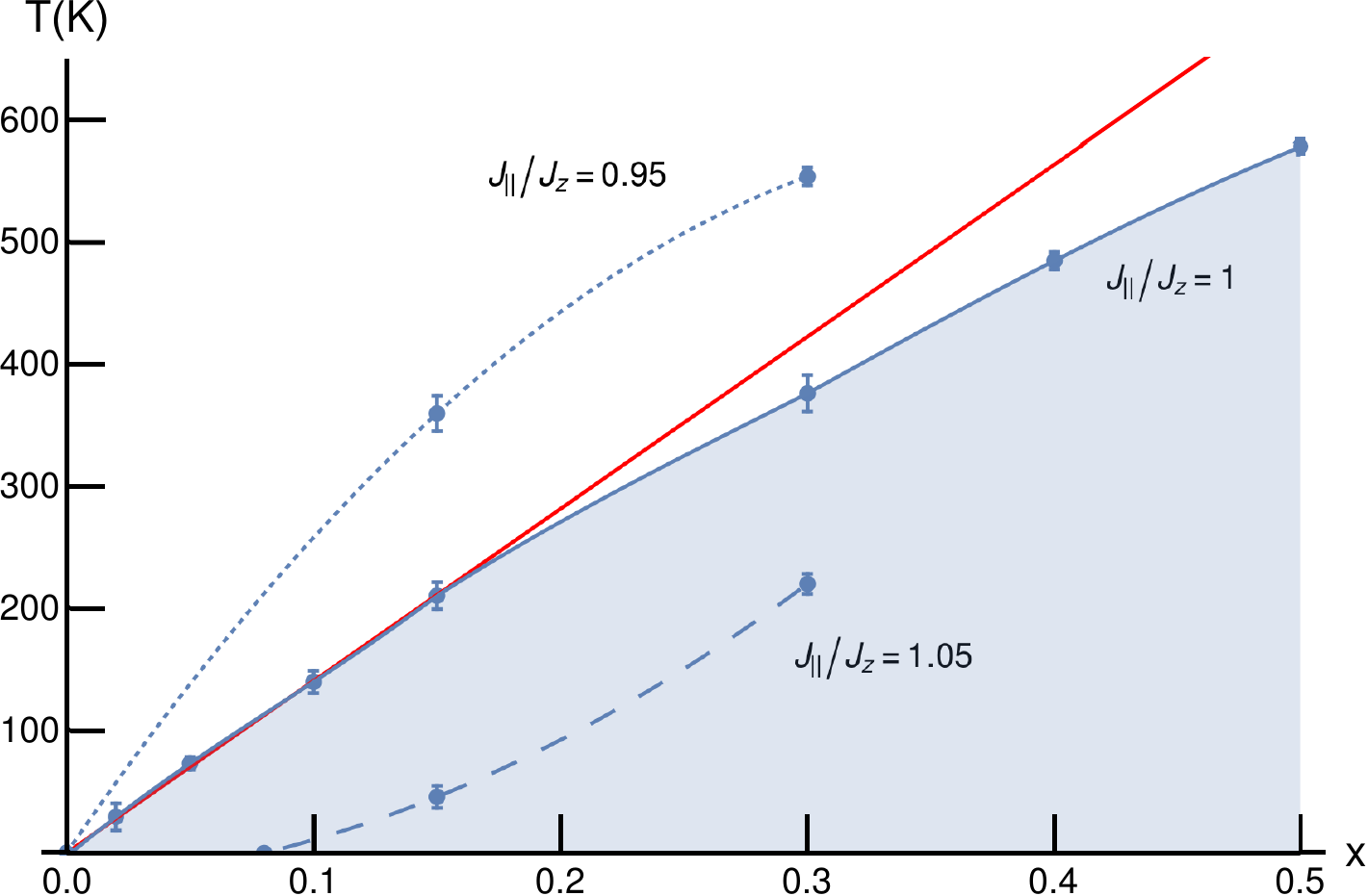}\hspace{3em}\includegraphics[width=0.9\columnwidth]{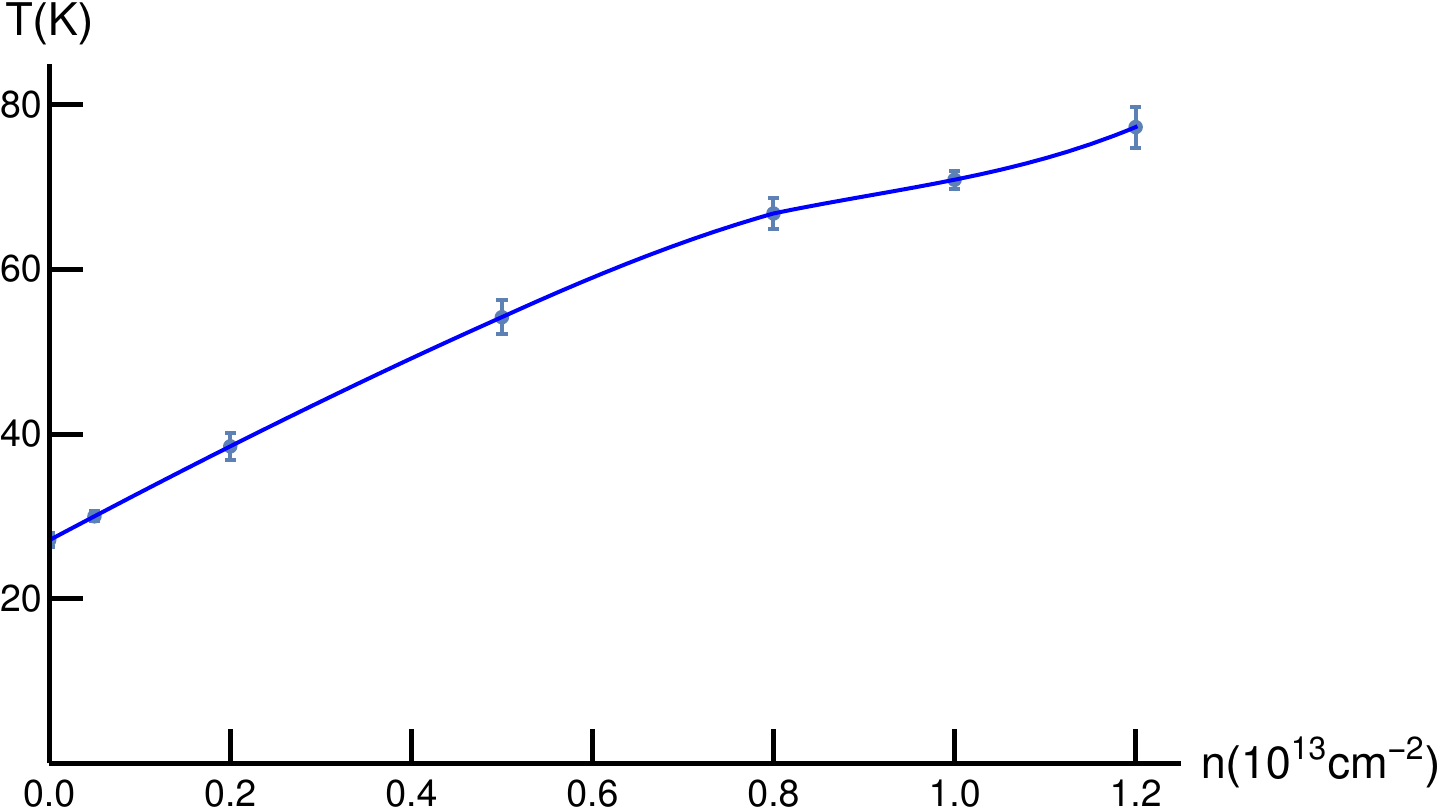}
\par\end{centering}
\caption{\label{fig:Tc_x}(left) Critical temperature for one-lattice ferromagnetism
at half filling. The full blue line represents the isotropic case,
which shows a linear behavior for small $x$, illustrated by the red
line. The dashed and dotted lines are for two different cases of anisotropy,
$J_{z}<J_{\parallel}$ and $J_{z}>J_{\parallel}$ , respectively.
In these cases, the critical temperature is greatly affected, not
only in terms of its value but also by becoming non linear. For $J_{z}<J_{\parallel}$
there is a critical concentration below which no ferromagnetic order
exists. (right) Critical temperature for one-lattice ferromagnetism
as a function of electron doping $n$. These results were obtained
for $JS=0.4t$ and $x=0.1$. There is a steady increase of $T_{C}$
as the electronic density increases within the experimentally relevant
regime. }
\end{figure*}

\section{Adatoms in a single sublattice\label{sec:single}}

\subsection{Ferromagnetic critical Temperature}

\label{subsec:1sublattTc}

\begin{figure*}
\noindent \centering{}\includegraphics[width=0.9\columnwidth]{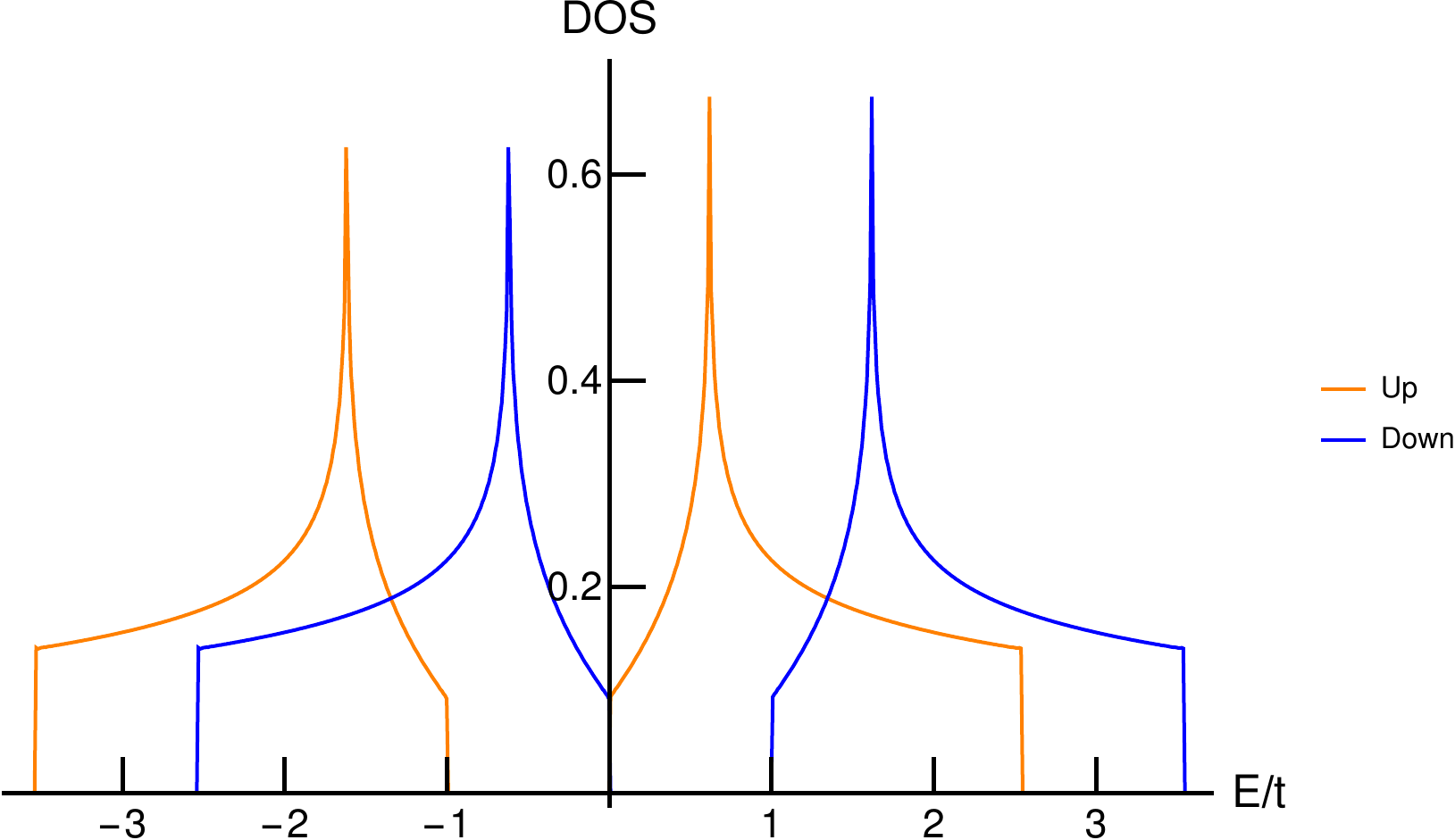}\hspace{3em}\includegraphics[width=0.9\columnwidth]{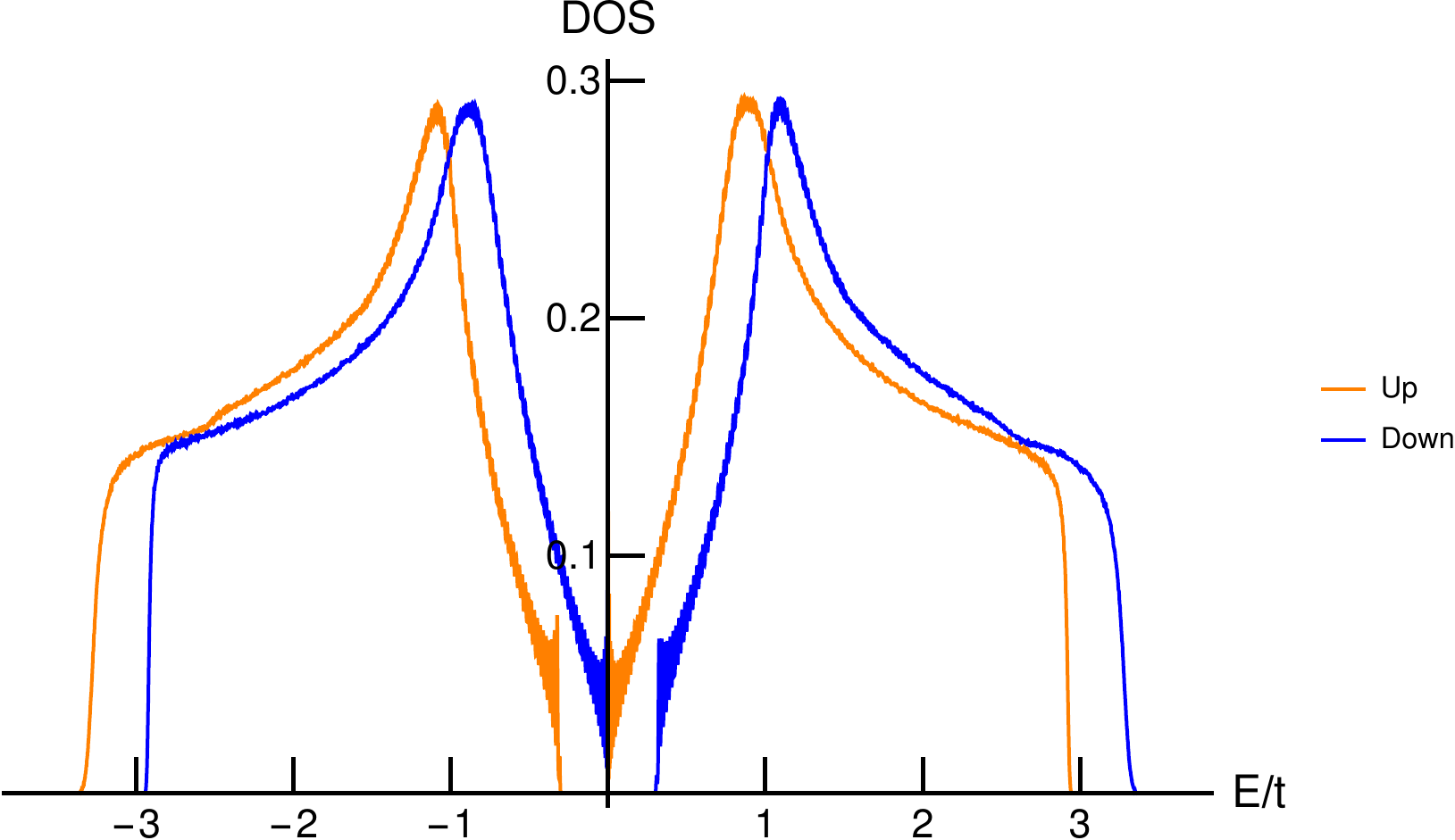}\caption{\label{fig:DOS_ferro_T0}DOS at $T=0$ for two different concentrations
of adatoms in a single sublattice: $x=0.5$~(left) and $x=0.15$~(right).
For $x=0.5$, we obtain the expected result with gaps of value $JS$
for each spin projection. At lower adatom concentrations, the states
become more uniformly distributed as the gaps decrease.}
\end{figure*}

When placed on only one of the sublattices, the classical spins tend
to align their orientations forming a ferromagnetic phase. In Fig.~\ref{fig:Tc_x}(left),
the critical temperature is shown for a half-filled graphene lattice.
In full, the isotropic case, $J_{\parallel}=J_{z}=J$, with $JS=t$
\citep{baskaran2007}, shows a linear behavior for low impurity concentrations,
with no signs of a critical concentration below which the ferromagnetic
order is lost. This agrees with approximate results for a system of
diluted spins interacting through a ferromagnetic, long ranged RKKY-like
interaction \citep{Szalowski2014}. For $x<15\%$, the temperature
dependence on $x$ is well described by the linear function $T_{C}\approx1400x\,\text{(K)}$.
The value of the coupling $JS$ obviously affects the critical temperature
and is dependent on the impurity species used, with higher values
resulting in higher $T_{C}$. 

As already argued, we expect the real system not to be completely
isotropic. We can easily change our model to encompass anisotropy,
by distinguishing between the inplane and perpendicular directions:
$J_{z}\neq J_{\parallel}$ (maintaining $J_{z}S=t$). We have found
that even a small amount of anisotropy ($5\%$) substantially alters
the $T_{C}$, as seen in Fig.~\ref{fig:Tc_x}(left). The dotted line
refers to the anisotropic case with $J_{z}>J_{\parallel}$, where
the critical temperature increases and becomes sub-linear. The case
$J_{z}<J_{\parallel}$ (dashed line) the behavior seems to be superlinear.
At low $x$, the suppression of $T_{C}$ is so strong that a critical
concentration shows up. In this case, at $T=0$, the system undergoes
a quantum phase transition as a function of $x$. In the very diluted
limit at low enough temperatures, Kondo physics could eventually take
over \citep{baskaran2007}. This effect is not considered here.

The methodology followed here allows also to study the case of graphene
away from half filling. In this situation, there should be more electrons
available to couple different impurities. Based on this argument we
can understand Fig.~\ref{fig:Tc_x}(right), where we show $T_{C}$
as a function of electron density, $n$, for a fixed concentration
of magnetic adatoms $x=0.1$. The ferromagnetic phase becomes more
stable as $n$ is increased. This holds even when the chemical potential
is located inside the pseudogap that shows up in this phase, to be
discussed below. Nevertheless, we expect the critical temperature
to eventually reach a maximum value for some electronic density, decreasing
towards zero after that. Our results show that this does not happen
for electronic densities experimentally relevant. 

The electronic density in graphene is easily changed by applying a
bias gate voltage. Interestingly, the result of Fig.~\ref{fig:Tc_x}(right)
provides proof of concept for a ferromagnetic transition tunable by
electrical means: for a fixed temperature, tuning the electronic density
through a gate voltage could induce ferromagnetic order on increasing
voltage and paramagnetic behavior when the voltage is decreased.

\begin{figure}
\begin{centering}
\includegraphics[width=0.95\columnwidth]{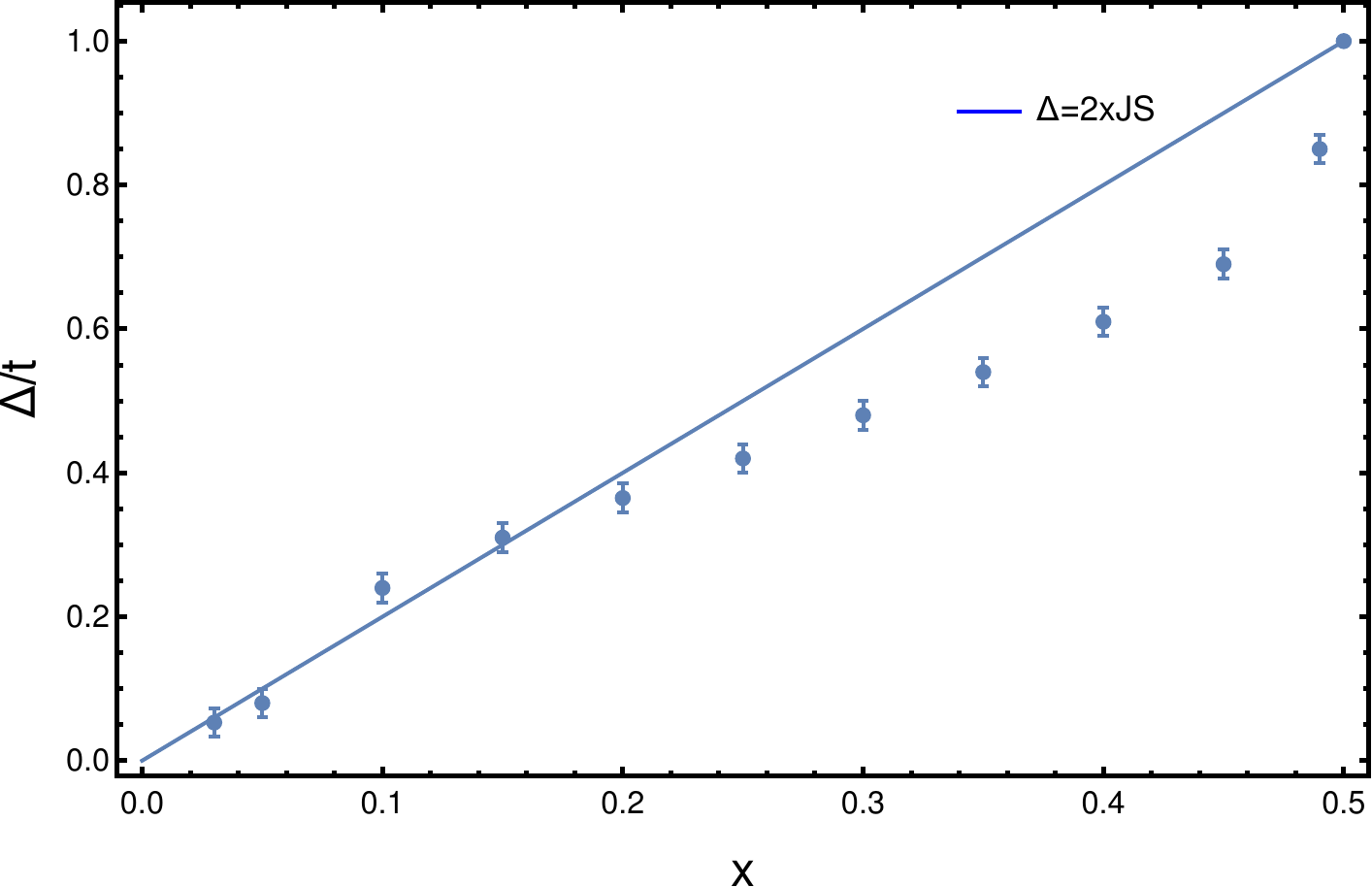}
\par\end{centering}
\caption{\label{fig:ferro_gap_comp}Spin resolved gap at $T=0$ for different
one sublattice adatom concentrations. The line represents the value
expected for a model where every site holds an adatom with smaller
coupling.}
\end{figure}

\subsection{Spectral properties }

\subsubsection{Single sublattice ferromagnetic phase \label{subsec:dos1LattFerro}}

In the ferromagnetic phase the two electron-spin projections are not
degenerate. When the impurities have a preferred direction, on average,
one electron-spin projection will gain energy through interactions
with the impurities, while the other will lose energy. At $T=0$,
the system displays only a spin-resolved gap that decreases as $x$
is lowered, as shown in Fig.~\ref{fig:DOS_ferro_T0}. Since there
is never an overlap between the gaps in each spin projected DOS, the
total DOS displays a pseudogap -- a region of energies where a depletion
of states is observed. 

Let us consider the full single sublattice coverage, $x=0.5$. In
this case the spin resolved gap is $JS$ and centered at $\pm JS/2$,
depending on the spin projection. In fact, the system is now translational
invariant and we can solve this case analytically in $k-$space \citep{daghofer-ssibmiig2010}.
The energy bands are given by

\begin{eqnarray}
E_{\text{\ensuremath{\mathbf{k}}},\uparrow} & = & \frac{1}{2}\left(-JS\pm\sqrt{J^{2}S^{2}+4|f_{\mathbf{k}}|^{2}}\right)\label{eq:E_band_up_ferro}
\end{eqnarray}
\begin{equation}
E_{\text{\ensuremath{\mathbf{k}}},\downarrow}=\frac{1}{2}\left(JS\pm\sqrt{J^{2}S^{2}+4|f_{\mathbf{k}}|^{2}}\right),\label{eq:E_band_down_ferro}
\end{equation}
where $f_{\mathbf{k}}=\sum_{\boldsymbol{\delta}}\exp\left(i\mathbf{k}\cdot\boldsymbol{\delta}\right)$.
We can use this to predict how the spin resolved gap decreases with
impurity concentration. As we lower $x$, we can think that the effect
of the coupling constant is diluted over every site of the sublattice,
so that the system retains its translational symmetry. Now we can
use the above equations with the modification $JS\rightarrow2xJS$.
In Fig.~\ref{fig:ferro_gap_comp} we see that the results follow
the predicted tendency, although the agreement is not complete. For
high impurity concentrations the gap is lower than the predicted by
this simple model. Both methods agree for concentrations below 20\%.

\begin{figure}
\noindent \begin{centering}
\includegraphics[width=0.9\columnwidth]{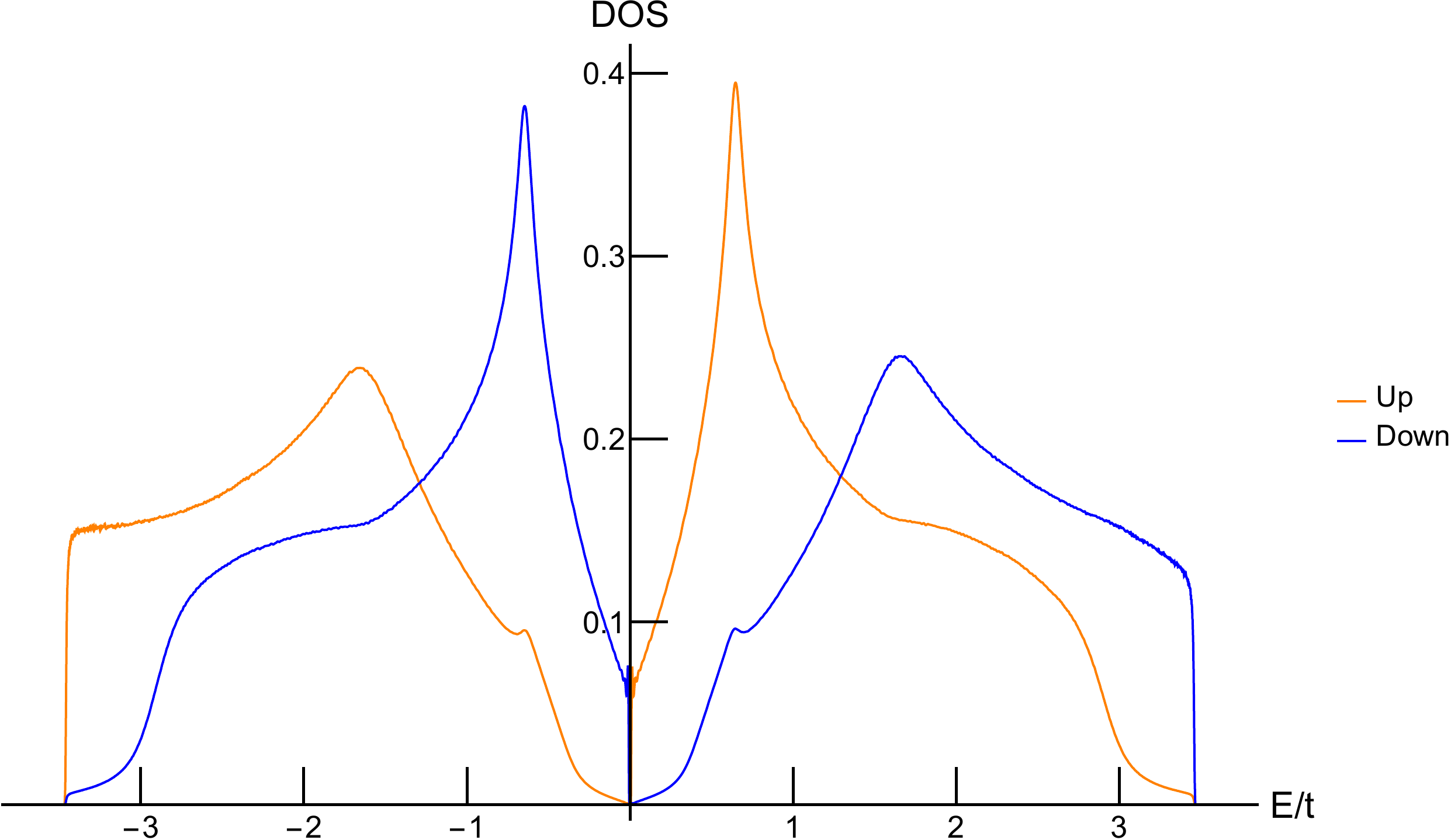}
\par\end{centering}
\noindent \begin{centering}
\includegraphics[width=0.9\columnwidth]{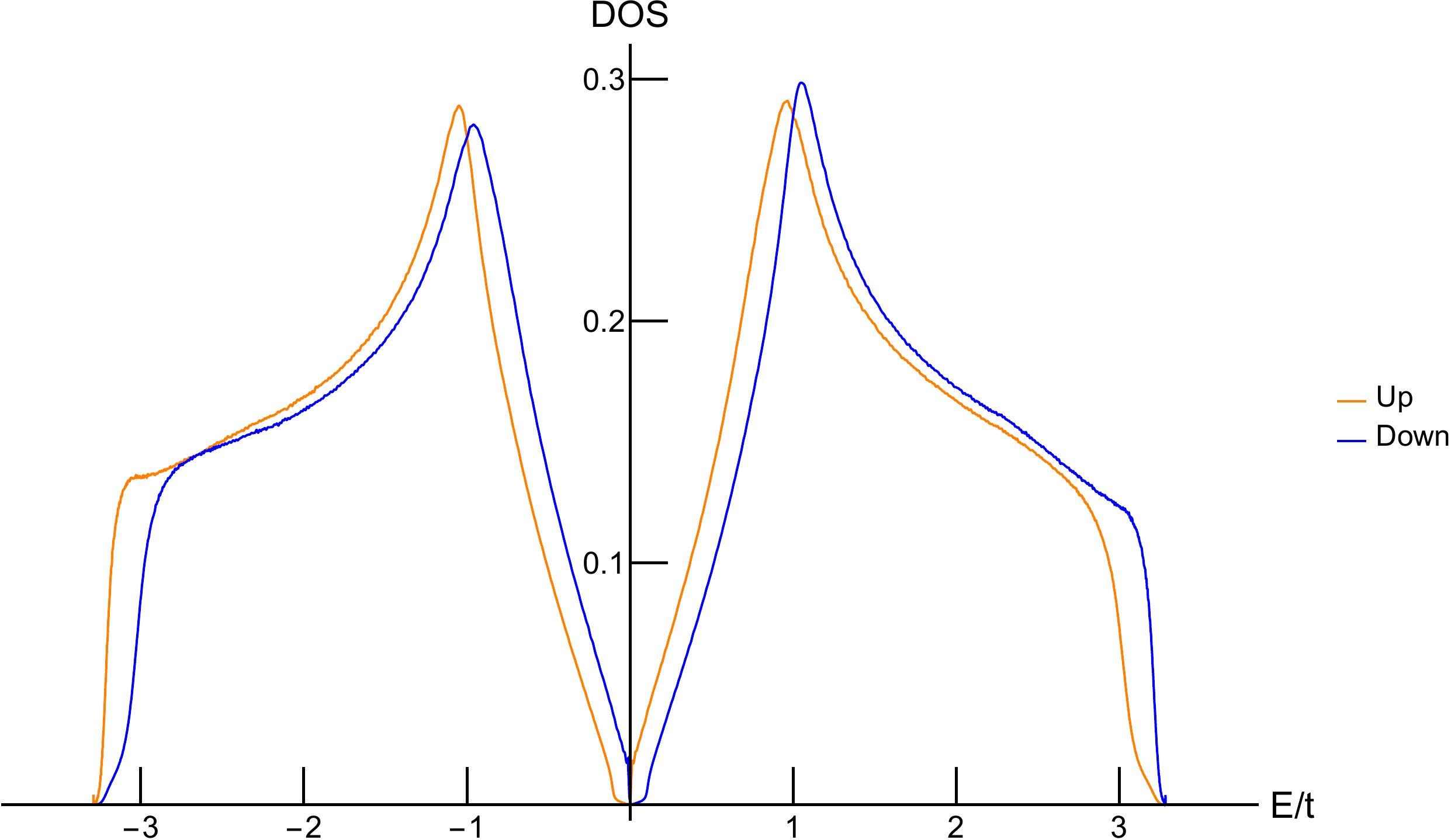}
\par\end{centering}
\noindent \centering{}\includegraphics[width=0.9\columnwidth]{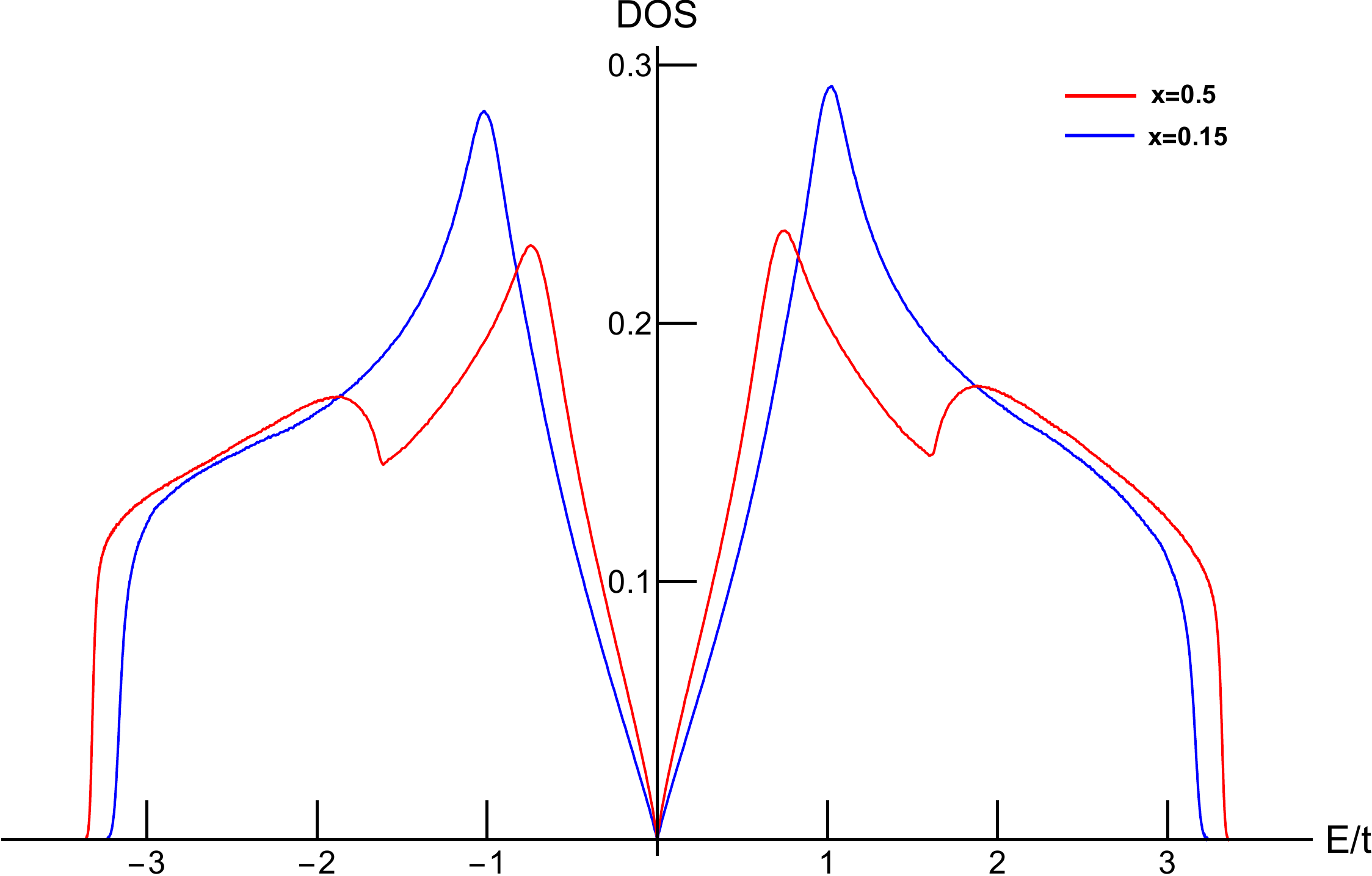}\caption{\label{fig:DOS_ferro_T<Tc}DOS at finite temperature, for different
impurity concentrations on only one sublattice: (top)~$T=0.72T_{C},\,x=0.5$;
(mid)~$T=0.82T_{C},\,x=0.15$. Temperature effects destroy the gaps
in the spin projected DOS, creating Lifshitz tails. For high values
of $x$ the spectrum for each spin is highly asymmetric.(bottom) DOS
in the single sublattice paramagnetic regime for $x=0.5$ (red) and
$x=0.15$ in (blue).}
\end{figure}

The finite temperature case is shown in Fig.~\ref{fig:DOS_ferro_T<Tc},
for $x=0.5$ (top) and $x=0.15$ (mid). We see that, for $0<T<T_{C}$,
the gaps of each spin projection are no longer present, since Lifshitz
tails effectively close the gaps. Both in this case and the $T=0$
case, there is always a region of energies where the electrons spin
polarization is not balanced. This is particularly true around the
zero energy, which means that any kind of electron doping will be
polarized. In fact, since the states that form the Lifshitz tails
are localized, we expect charge carriers to be 100\% spin polarized
in the pseudogap region. It is also worth mentioning the asymmetry
observed for each spin projected DOS, especially evident at high concentrations.
The main conclusion here is that, even though the spin resolved gaps
vanish at finite temperature, there is still an imbalance between
the two spin projections so that the pseudogap is still present. Ultimately,
this feature is responsible for the energy gain in the ferromagnetic
phase.

\subsubsection{Single sublattice paramagnetism}

In the paramagnetic regime, since the spins of the impurities are
randomly oriented, there is no overall energy gain or loss for either
electron-spin projection. Therefore, the system is now spin degenerate.
In Fig.~\ref{fig:DOS_ferro_T<Tc}(bottom) we see that, similarly
to the clean graphene case, the DOS vanishes linearly at $E=0$. In
the clean limit, the DOS at low energies behaves as $\rho(E)=|E|/(hv)^{2}$,
where $v$ is the Fermi velocity. An effective electron velocity $\tilde{v}$
may then be defined by linearly fitting the low energy numerical result
in Fig.~\ref{fig:DOS_ferro_T<Tc}(bottom). We obtain, for $x=0.5$
a velocity of $\tilde{v}_{x=0.5}=0.66v$, and for $x=0.15$ the velocity
$\tilde{v}_{x=0.15}=0.85v$. For $x=0.05$ we get a velocity $\tilde{v}_{x=0.05}=0.94v$,
so we can actually infer a relation for the electronic velocity in
this regime of low concentrations, $\tilde{v}_{x}\approx(1-x)v$.
So, electrons around zero energy move slower when we include adatoms
with their spins randomly orientated. We must keep in mind, however,
that we are not taking into account the scattering of electrons in
the adatoms \citep{Irmer2018}, which should further slow the movement
of electrons on the lattice.

\begin{figure}
\begin{centering}
\includegraphics[width=0.9\columnwidth]{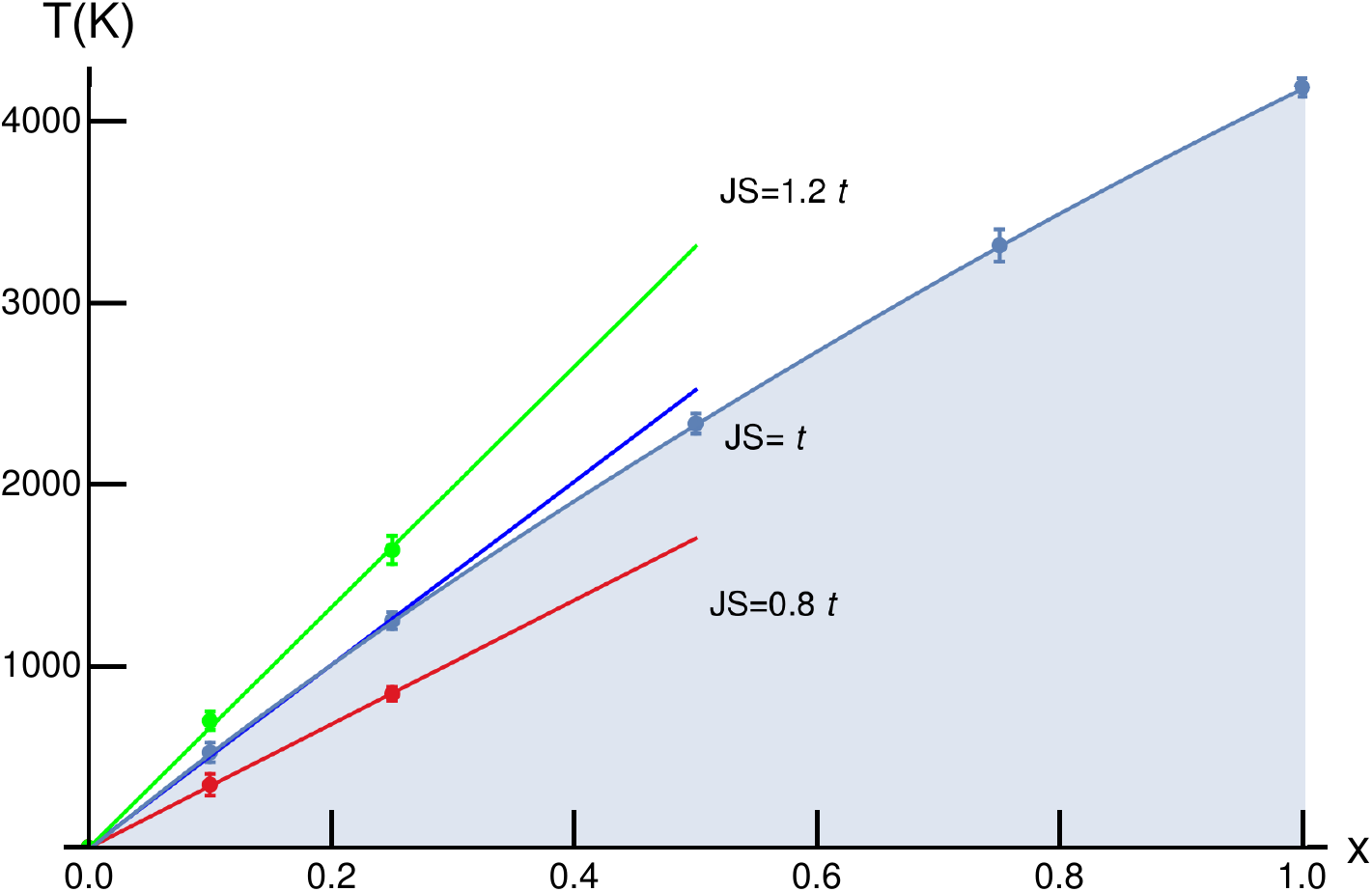}
\par\end{centering}
\caption{\label{fig:DOS_ferro_para-TcAF}Critical temperatures for antiferromagnetism
when both lattices are randomly occupied, on equal amount. The linear
dependence for low $x$ is apparent.}
\end{figure}

\section{Adatoms in both sublattices\label{sec:both}}

\subsection{Critical Temperature}

When magnetic adatoms are allowed to sit on both sublattices an antiferromagnetic
regime is favored. In Fig.~\ref{fig:DOS_ferro_para-TcAF} we show
the critical temperature below which an antiferromagnetic long range
order starts to develop at mean field level, as a function of the
impurity concentration $x$ for different values of coupling parameter
$JS$. This case displays similar features to the single sublattice
ferromagnetism, also predicting the absence of a critical concentration.
The values of $T_{C}$, however, are one order of magnitude greater
than the ferromagnetic critical temperature studied in Sec.~\ref{subsec:1sublattTc}.
For $JS=t$, the relation $T_{C}\approx5000x\,\text{(K)}$ provides
a good description of the linear behavior.

It is worth noting that Quantum Monte Carlo results for RKKY-like
models lead to a much higher critical temperature for the antiferromagnetic
case when the oscillatory component is not taken into account \citep{fabritius2010}.
So, MF seems to be a satisfactory approximation in this case. The
colored lines in Fig.~\ref{fig:DOS_ferro_para-TcAF} show the $T_{C}$
follows the expected behavior with the change of the coupling constant
$JS$: higher critical temperature for higher coupling, lower critical
temperature for lower coupling. We have verified that $T_{C}$ has
a quadratic dependence on $J$ (not shown), as found in Ref.~\citep{rappoport2011magnetic}.

\subsection{Spectral properties}

As a general feature of antiferromagnetic order, we obtain that the
two electron-spin projections are always degenerate. This can be understood
as follows. Since impurities in different sublattices tend to align
in opposite directions, electrons with one spin projection will gain
energy in, say, sublattice A and lose energy in sublattice B. The
same happens for the electrons with the other spin projection (with
the roles of sublattices A and B interchanged). So, the two spin projections
are equivalent, and thus we obtain a degenerate spin-resolved DOS.
Throughout this section we consider $JS=t$.

\subsubsection{Full coverage antiferromagnetism}

\begin{figure*}
\noindent \centering{}\includegraphics[width=0.9\columnwidth]{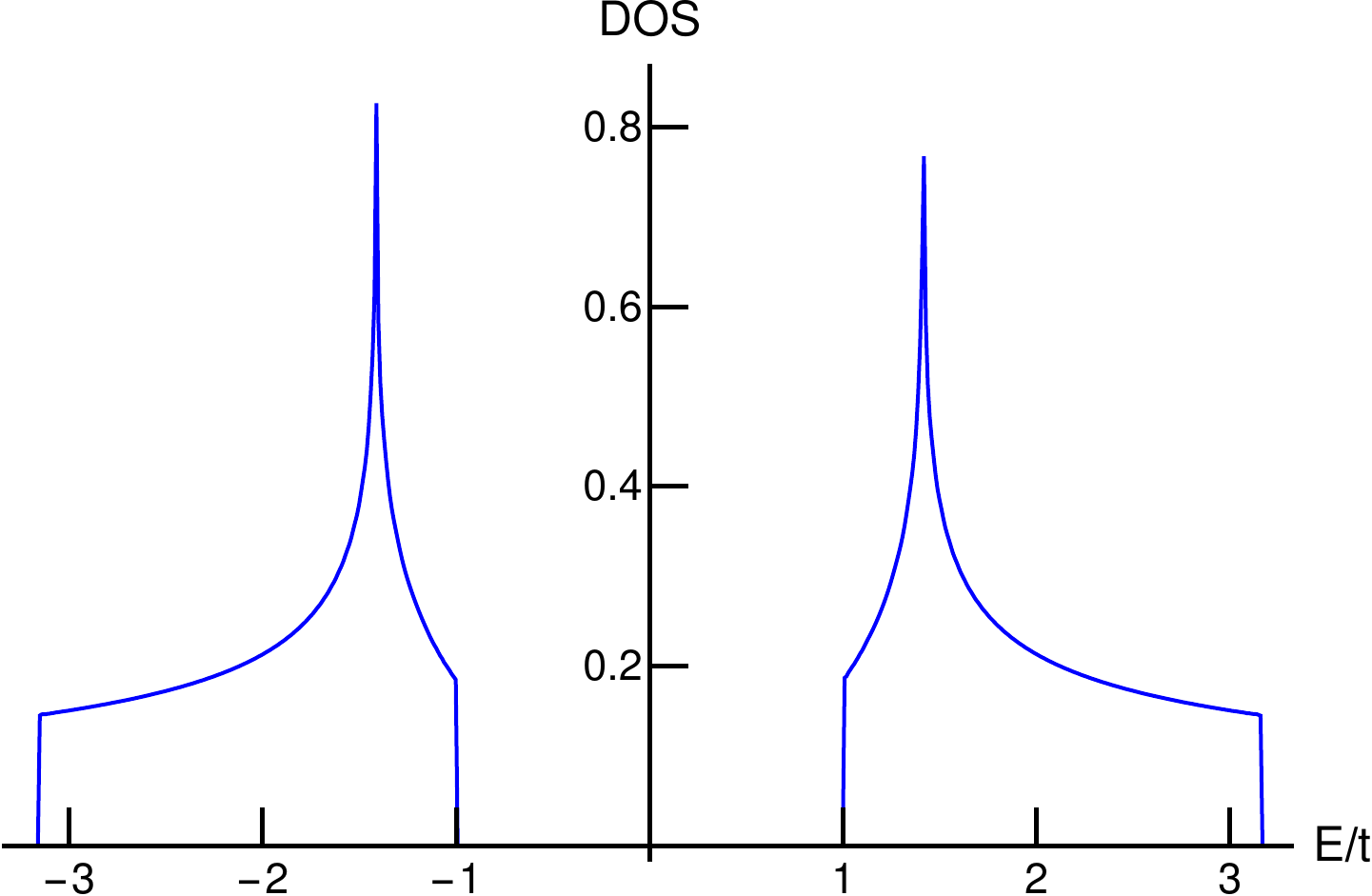}\hspace{3em}\includegraphics[width=0.9\columnwidth]{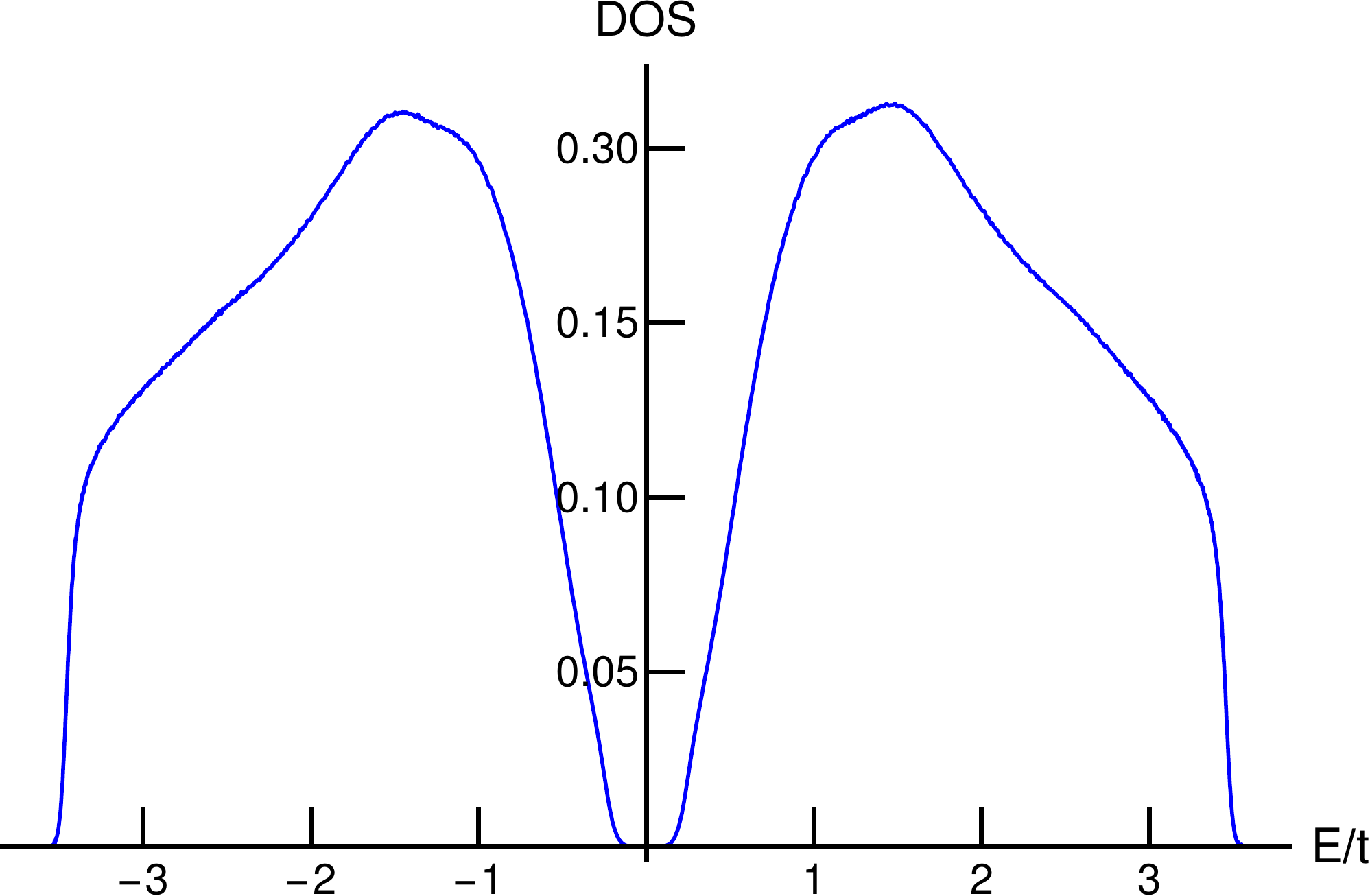}\caption{\label{fig:DOS_anti_100}DOS for the full coverage antiferromagnetic
phase at $T=0$~(left), when no thermal disorder exists, and $T=0.73T_{C}$~(right).
Notice how singularities and band edges get smoothed out due to thermally
induced disorder.}
\end{figure*}

In order to gain insight into the obtained results, let us consider
the fully covered, $x=1$ case. At $T=0$ we obtain a gap centered
at zero energy, with value $2JS$, as shown in Fig.~\ref{fig:DOS_anti_100}(left).
Note that at $T=0$, when all thermal fluctuations are suppressed
(i.e., at mean field level, all configurations of classical spins
which deviate from perfect Néel order are inaccessible), we recover
translational invariance due to the perfect long range antiferromagnetic
order of the magnetic adatoms. In this limit we can solve the problem
in $k$-space, as we did ferromagnetic ordering in Sec.~\ref{subsec:dos1LattFerro},
in oder to obtain the spectrum and thus the DOS \citep{daghofer-ssibmiig2010}.

If we increase the temperature, but keep the full coverage $x=1$
case, we can isolate the effect of spin orientation disorder. The
gap decreases significantly, with Lifshitz tails smoothing its edges,
and the states get much more uniformly distributed along the whole
energy spectrum, erasing the Van-Hove singularities. This effect is
illustrated in Fig.~\ref{fig:DOS_anti_100}(right). So, in contrast
with the single sublattice ferromagnetic phase, the system opens an
energy gap that survives at finite temperatures. 

\subsubsection{Two-sublattices, partial coverage antiferromagnetism}

\begin{figure}
\begin{centering}
\includegraphics[width=0.9\columnwidth]{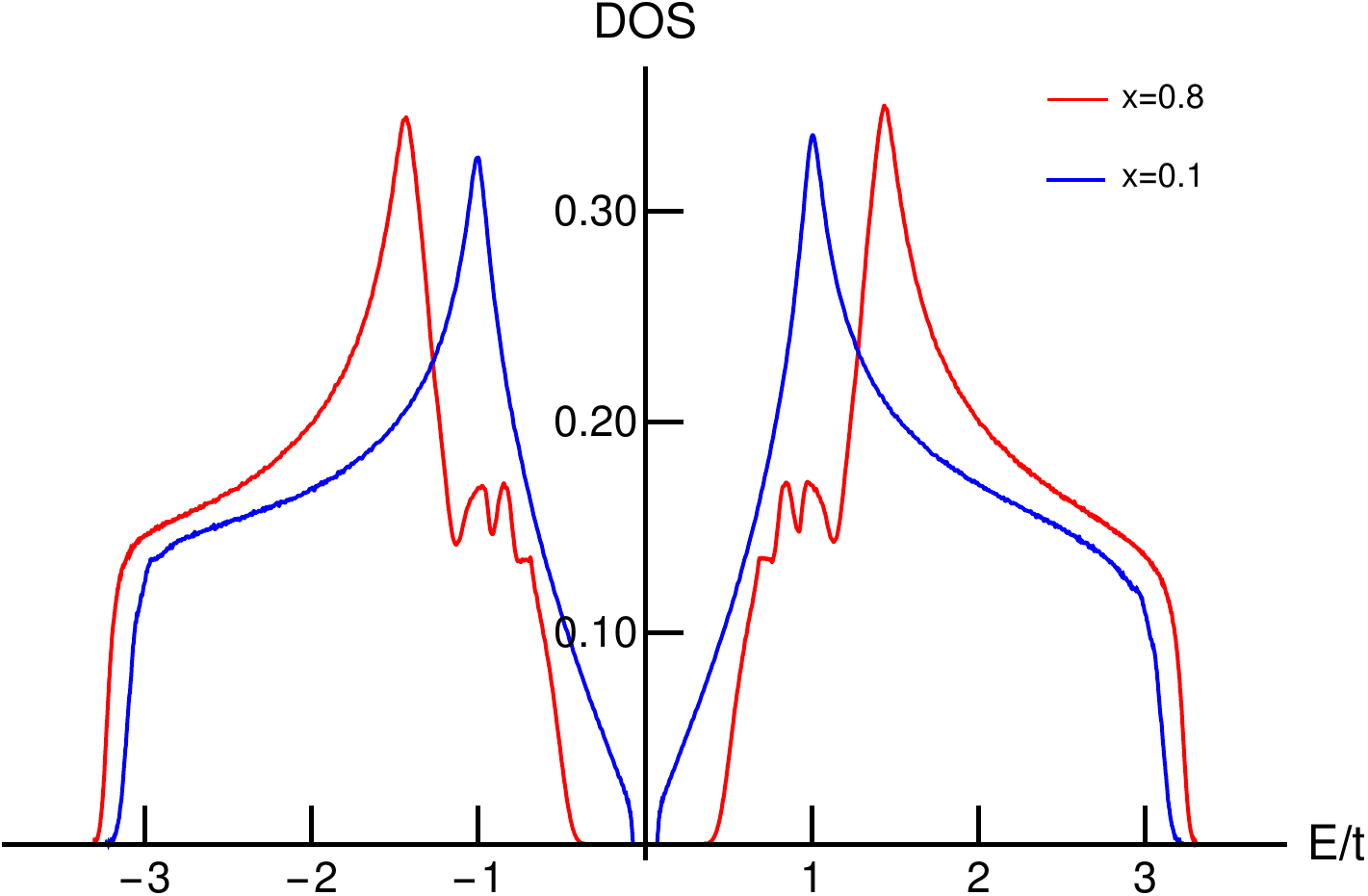}\vspace{0.5cm}
\par\end{centering}
\begin{centering}
\includegraphics[width=0.9\columnwidth]{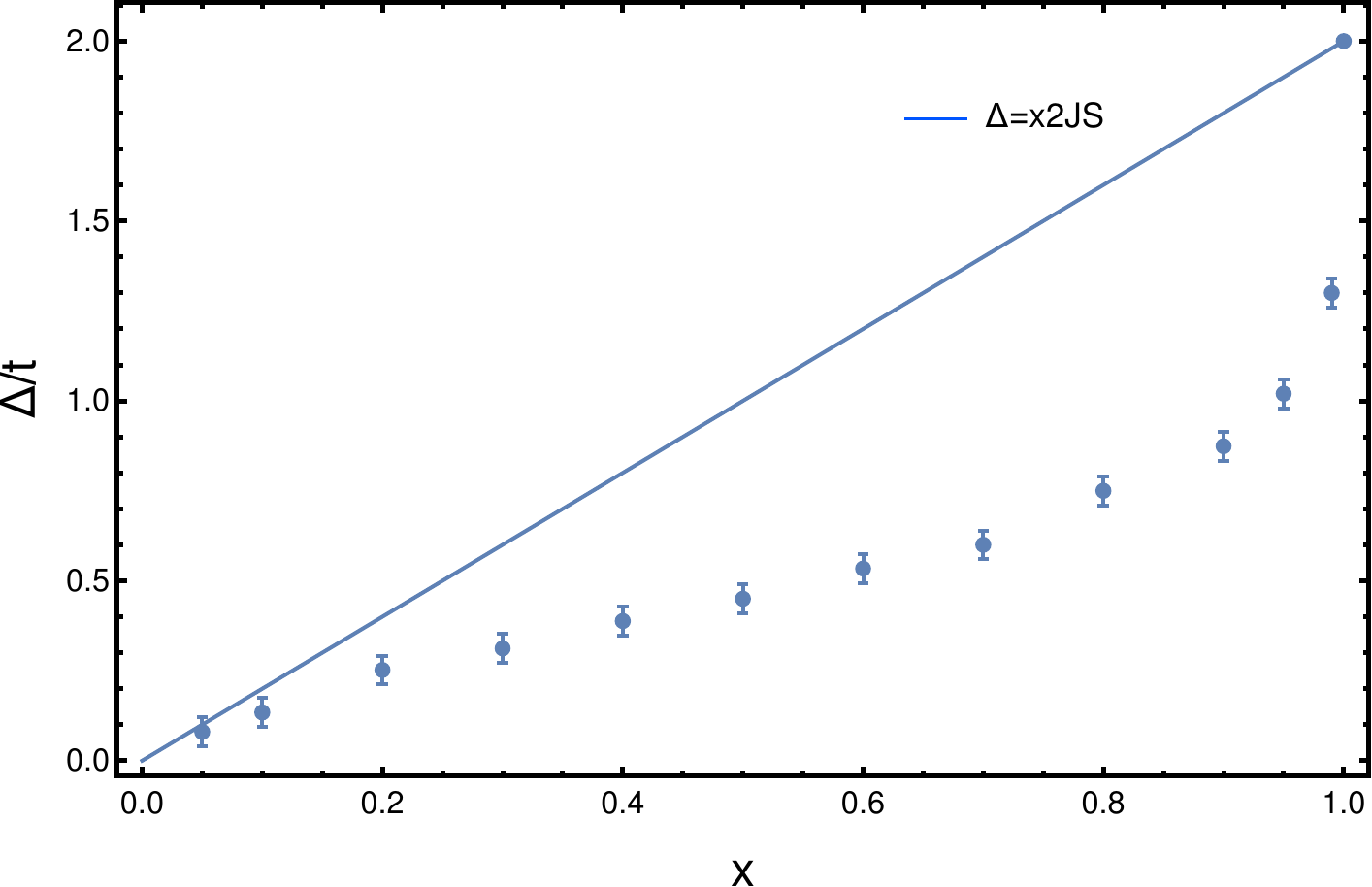}\vspace{0.5cm}
\par\end{centering}
\begin{centering}
\includegraphics[width=0.9\columnwidth]{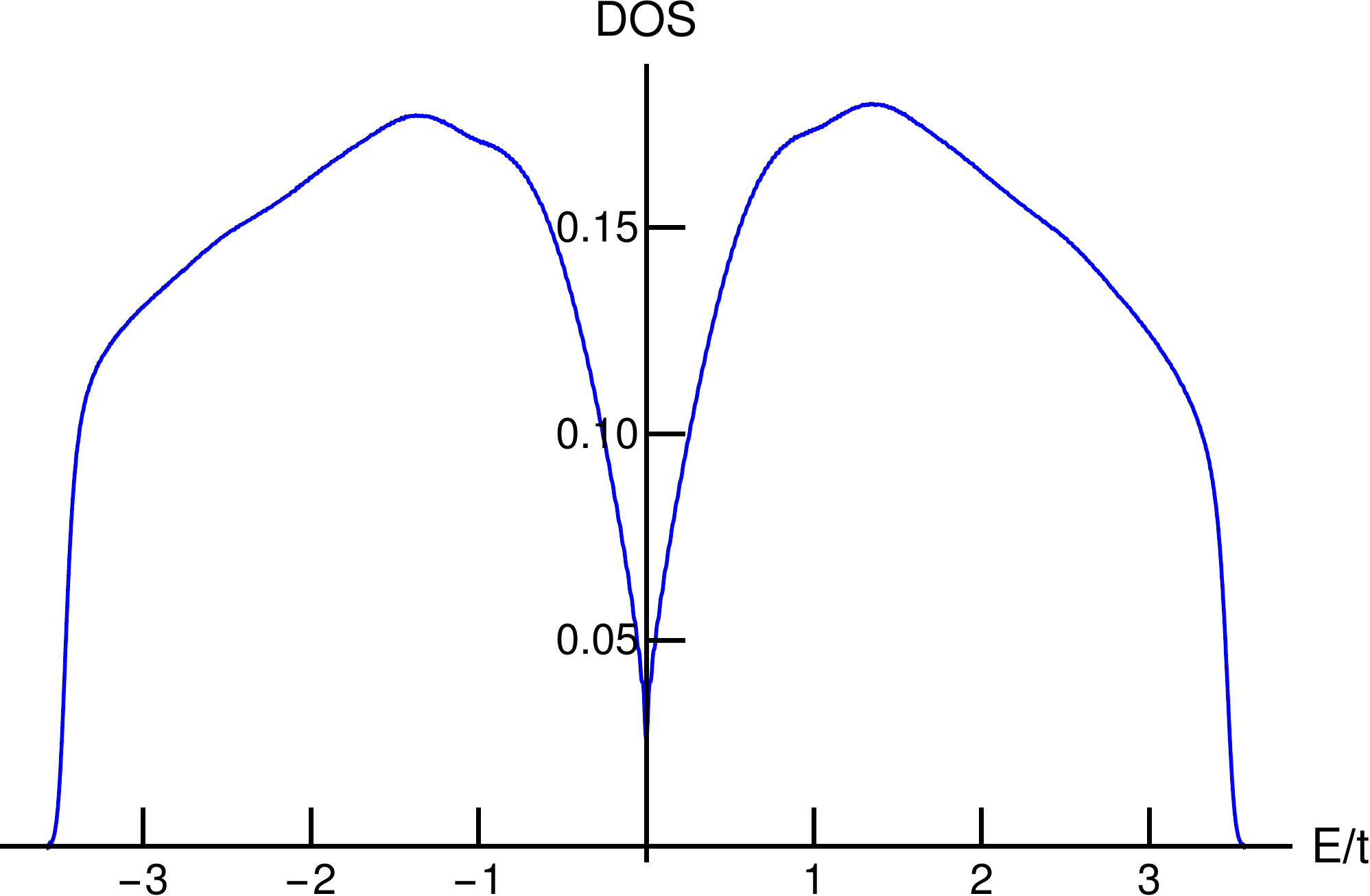}
\par\end{centering}
\caption{\label{fig:DOS_anti_T0-Gap}(top) DOS in the antiferromagnetic regime
for $x=0.8$ (red) and $x=0.1$ (blue), at zero temperature. The energy
gap decreases as we lower the value of $x$. At $x=0.8$ we can see
the peak-like structures near the gap edges. (mid) Energy gap at $T=0$
for different concentrations. The line represents the value expected
using a model where every site is occupied but with a weakened coupling.
(bottom)DOS in the paramagnetic regime for $x=0.75$ with impurities
on both sublattices. The DOS is finite at $E=0$}
\end{figure}

Now we keep $T=0$, when perfect antiferromagnetic order is present,
and study the effect of adatom position disorder by taking $x<1$.
In Fig.~\ref{fig:DOS_anti_T0-Gap}(top) the DOS for $x=0.8$ and
$x=0.1$ is shown. We see immediately that the energy gap decreases,
although it is present at any concentration. For high impurity concentrations
there are is a reconstruction of the DOS right next to the gap. The
associated structures merge with the Van Hove singularities at $x\approx0.5$.
Increasing the temperature also destroys these features.

To understand the origin of these states let us go back to the full
coverage case at $T=0$. There, the entire lattice is covered with
impurities whose spins are oriented either up or down, depending on
the sublattice they are located at. It is this full coverage that
is responsible for the gap between $-JS$ and $JS$. If we remove
a few impurities from the lattice, the gap should suffer only a slight
perturbation. Additionally, these missing impurities create local
states with midgap energies that start off as delta-like peaks in
the DOS. Lowering the concentration broadens these sates and they
eventually get included in the bands as the gap decreases. These impurity
states are easily seen in the DOS, obtained with the recursion method,
for example at $x\approx0.99$ (not shown).

We can make an analysis similar to the that of Sec.~\ref{subsec:dos1LattFerro},
spreading the effect of impurities over all sites, so that the obtained
gap is given by $\Delta=x2JS$. In Fig.~\ref{fig:DOS_anti_T0-Gap}(mid)
we can see how this model matches up with the results. As soon as
we leave the full coverage case there is a steep decrease of gap relative
to the $x2JS$ line. This is due to the states already mentioned that
are created inside the gap, effectively reducing it. As the concentration
decreases, the two approaches start to yield similar results. This
is especially evident for concentrations under 10\%. The energy gap
present ensures the higher stability of the antiferromagnetic regime
compared to the single sublattice ferromagnetic phase, surviving both
adatom positioning and adatom-spin orientation disorders.

\subsubsection{Two-sublattices, paramagnetic phase}

Above the critical temperature, the impurity spins are randomly oriented.
The fact that now we have impurities on both sublattices, leads to
a finite DOS at $E=0$, as shown in Fig.~\ref{fig:DOS_anti_T0-Gap}(bottom).
This observation contrasts with the result obtained for the single
sublattice case, shown in Fig.~\ref{fig:DOS_ferro_T<Tc}(bottom).
As we lower $x$, the DOS in that point decreases. Below $x\approx0.10$,
we no longer have the required numerical resolution to conclude whether
the DOS is still finite or not. The transition from finite DOS at
$E=0$ in the paramagnetic phase to an energy gap in the antiferromagnetic
one seems to be abrupt, since for a temperature $T=0.95T_{C}$ there
is already a gap. So, most likely the gap is formed as soon as there
is a preferred direction for the impurity spins, regardless of the
value of the DOS at zero energy during the paramagnetic phase.


\section{Comparison with experiment\label{sec:Comparison-with-experiment}}

In this section we bring our model closer to experimental values to
see if it can be used to understand some of the results obtained by
Hwang et al. in their work concerning sulfur decorated graphene \citep{lanzara2016}.
They report a sulfur (S) concentration of $x\approx10\%$ and perform
ARPES and magnetotransport measurements. Their main findings can be
summarized as follows: temperature dependent depletion of states at
the Fermi energy; magnetoresistance compatible with magnetic hysteresis.

We note that S atoms are not expected to be magnetic. However, in
the case of Ref.~\citep{lanzara2016}, there is a finite charge transfer
from graphene to the S atoms that is measured experimentally and which,
according to DFT calculations~\citep{lanzara2016}, is responsible
for the formation of a magnetic moment of $0.63\,\mu_{B}$ per S atom.
Moreover, according to the same DFT calculations \citep{lanzara2016},
a possible position for the S atoms is to occur underneath graphene,
between the top graphene layer and the buffer layer (a carbon layer
with the same structure of graphene but without the $\pi$-bands due
to strong hybridization with the SiC substrate). When the buffer layer
and graphene are Bernal stacked, the two sublattices of the top graphene
layer are no longer equivalent: one sublattice has buffer layer C
atoms below, while the other sublattice occurs at the buffer layer
hollow position where the S atoms sit (see Ref. \citep{lanzara2016}). 

Under the setup just presented, our model for one sublattice ferromagnetism
may be seen as an adequate starting point. This model explains qualitatively
the two main experimental observations of Ref.~\citep{lanzara2016}.
Regarding the magnetoresistance, and assuming that a ferromagnetic
state develops as predicted by the present theory, the resistance
should be maximized when the applied magnetic field reaches the value
of the coercive field. In this situation the misalignment between
the magnetization of different magnetic domains is maximum (so that
the magnetization is zero), so electron scattering by impurity spins
is also maximum. This explains the two peaks observed in the resistivity
in Fig.~4D of Ref.~\citep{lanzara2016} at the two opposite values
of the coercive field. This mechanism is particularly relevant in
the pseudogap region (see Fig.~\ref{fig:DOS_ferro_T<Tc}), where
it is expected that charge carriers at the Fermi level have a high
degree of spin polarization in the direction of the local magnetization.
As is shown below, the pseudogap is the relevant regime in the experiment
of Ref.~\citep{lanzara2016}.

\begin{figure*}
\begin{centering}
\includegraphics[width=0.9\columnwidth]{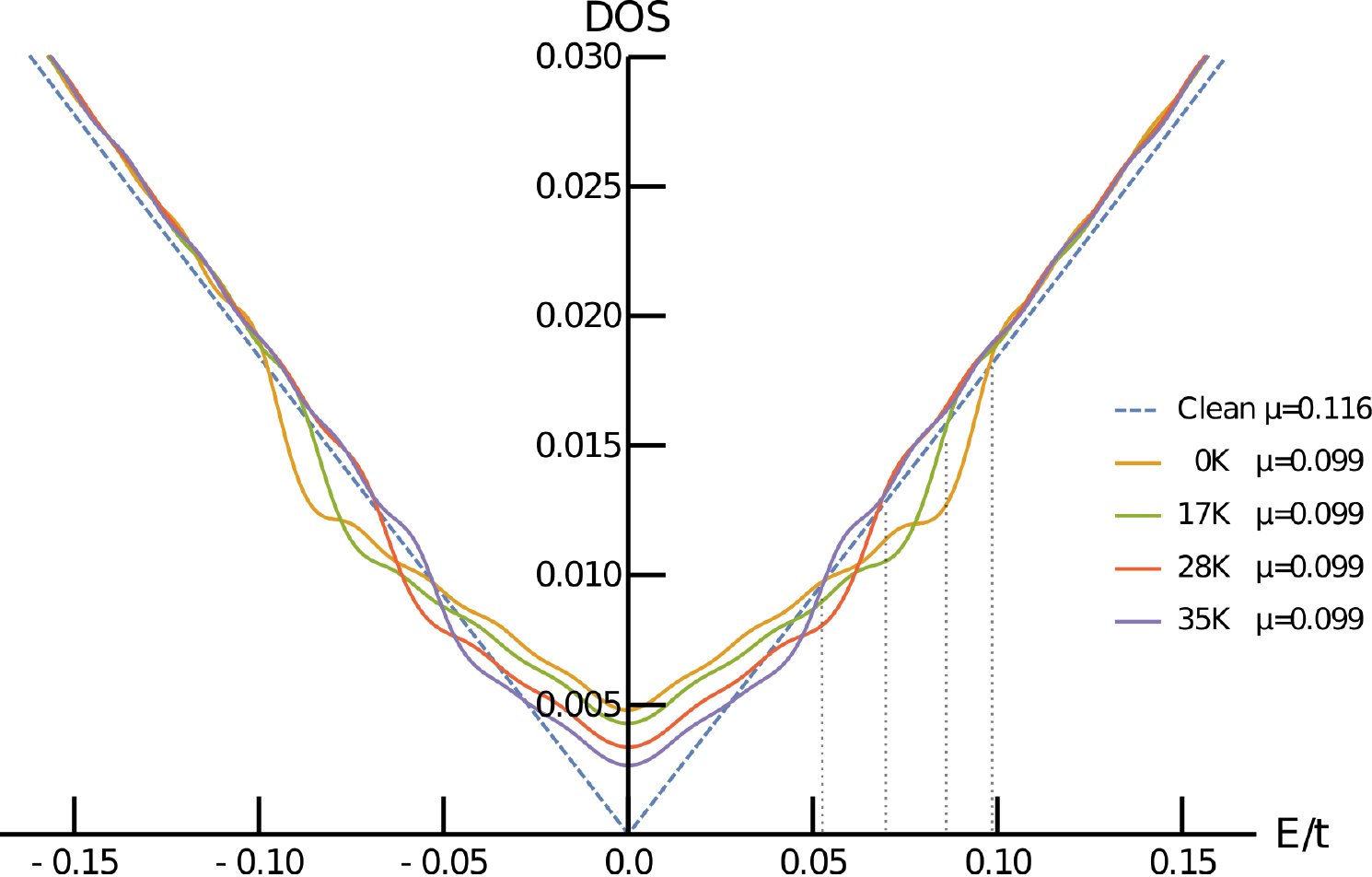}\hspace{3em}\includegraphics[width=0.95\columnwidth]{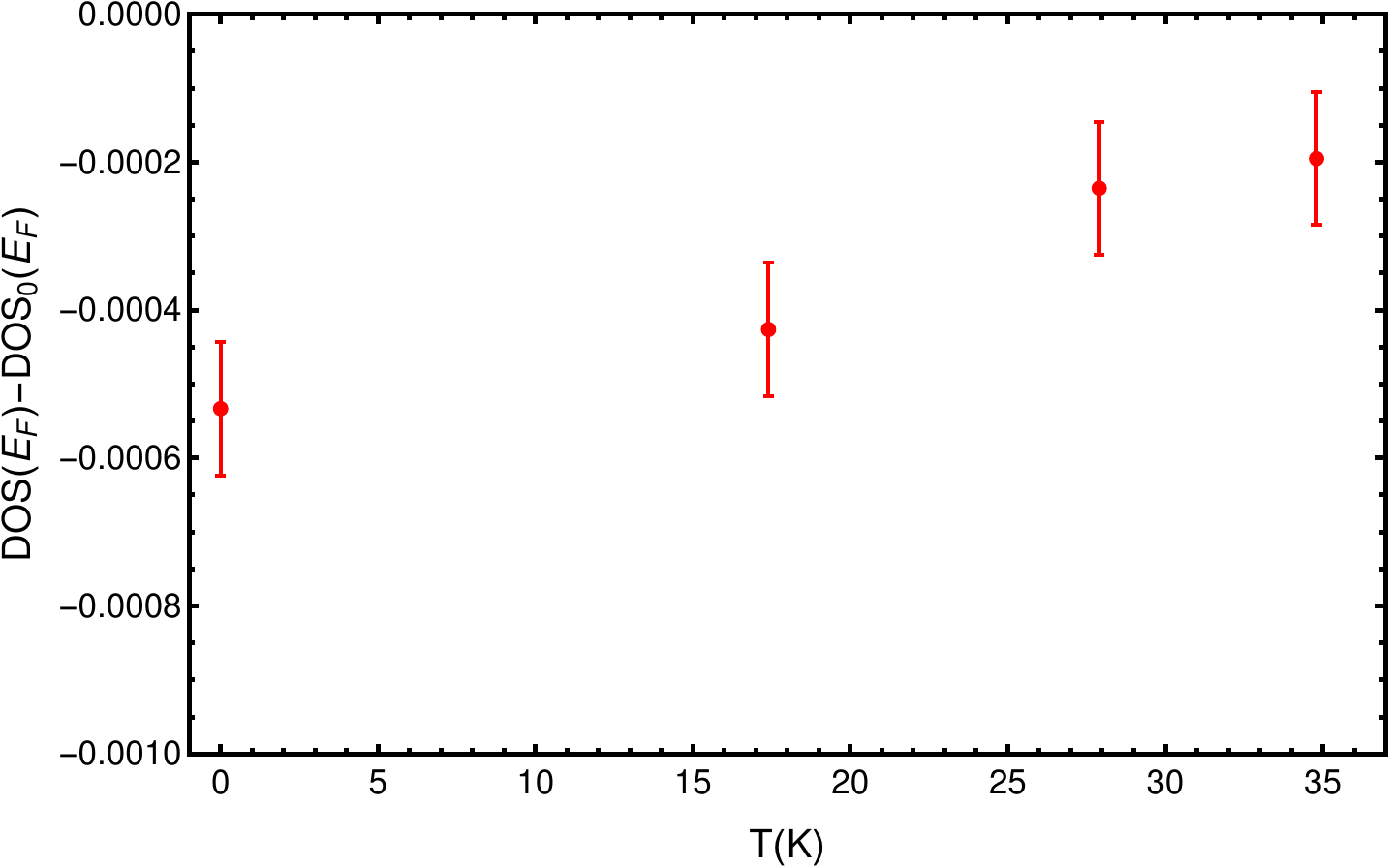}
\par\end{centering}
\caption{\label{fig:DOS_depletion} (left) Close up of the DOS around the pseudogap
for several different temperatures in the case of single sublattice
ferromagnetism for an impurity concentration $x=10\%$. The chemical
potential was determined for each temperature in order to keep the
electronic density fixed at $n=0.75\times10^{13}\,\text{cm}^{-2}$(right)
Difference between the DOS at the Fermi level for the sulfur doped
graphene and the clean graphene at different temperatures. According
to Ref.~\citep{lanzara2016}, the charge carrier density in the clean
case is $n=0.9\times10^{13}\,\mbox{cm}^{-2}$, while in the sulfur
decorated case it $n=0.75\times10^{13}$cm$^{-2}$. We used these
values to fix the Fermi energy in each case.}
\end{figure*}

We now turn to the depletion of states seen at the Fermi level in
Ref.~\citep{lanzara2016}. This result was interpreted as a signature
of the opening of a gap at the Fermi level. This interpretation is
hard to justify because of the lack of a nesting vector in the system,
assuming the S atoms are randomly distributed, which is the relevant
situation experimentally. In the present theory, the system does not
open a true gap, but a pseudogap at the Dirac point, which could also
explain the depletion of states near the Fermi level that is observed
experimentally. In Fig.~\ref{fig:DOS_depletion}(left) we show the
evolution of the pseudogap with the temperature. This pseudogap is
a consequence of the impurities, which create a gap in the spectrum
for each spin direction, one at positive energies and another at negative
energies (as shown in Figs.~\ref{fig:DOS_ferro_T0} and~\ref{fig:DOS_ferro_T<Tc}),
so there is a point where one of the spin resolved DOS is highly suppressed,
creating this depletion of states relative to the clean graphene layer
{[}shown as a dashed line in Fig.~\ref{fig:DOS_depletion}(left){]}.
The energy below which the depletion is most pronounced, signaled
in Fig.~\ref{fig:DOS_depletion}(left) by the vertical dotted lines
(roughly half the value of the pseudogap), becomes closer and closer
to zero energy as we increase the temperature. This is the effect
of disorder destroying the spin resolved gaps. On the other hand,
on approaching the Dirac point, we see a region where the DOS is enhanced
compared to the pristine case. In the latter case the DOS vanishes
linearly whereas with impurities there is always a contribution from
one of the spin projections.

In order to make a closer comparison with the experimental results
of Ref.~\citep{lanzara2016} regarding the depletion of states around
the Fermi energy, we plot in Fig.~\ref{fig:DOS_depletion} the difference
between the DOS at the Fermi level for the S decorated case and the
clean case. We determine the Fermi level by adjusting the charge carrier
density to the reported values in Ref.~\citep{lanzara2016}. For
the clean graphene, they report $n=0.9\times10^{13}\,\mbox{cm}^{-2}$,
from which we get $\mu\simeq0.116t$. Once the S impurities are added,
Hwang \emph{et al.} measure a change in the Fermi wave vector due
to a charge transfer between S-atoms and the graphene system. Because
of this charge transfer, the charge carrier density in graphene is
lowered to $n=0.75\times10^{13}\,\mbox{cm}^{-2}$. The obtained chemical
potential is $\mu\simeq0.099t$, with no significant change with temperature
as indicated in Fig \ref{fig:DOS_depletion}(left). The theory has
a single adjustable parameter, the value of the coupling $JS$, which
is fixed in order to reproduce the $T_{C}\sim30\,\text{K}$ reported
in Ref.~\citep{lanzara2016} at $x=10\%$ {[}a similar value has
been used to produce Fig.~\ref{fig:Tc_x}(right){]}. The negative
values obtained for the DOS difference shown in Fig.~\ref{fig:DOS_depletion}
is indicative of a depletion of states with respect to the clean case,
consequence of the shifting of the Fermi level towards lower energies.
This observation agrees with the experimental finding. Moreover, we
also see a temperature dependent depletion of states, where higher
values of depletion occur as the temperature is lowered. This happens
because the Fermi level is located at the edge of the pseudogap, whereas
for higher temperatures it is already outside that region. This finding
is again in agreement with the experimental result. In the present
theory such depletion has no relation with a gap opening at the Fermi
level.


\section{Conclusions\label{sec:Conclusions}}

We have studied the magnetic phase diagram of graphene decorated with
magnetic adatoms located at the top position of the graphene lattice.
Using a phenomenological $s$-$d$ model to couple the impurities
with the underlying graphene electrons, and working with classical
impurity-spins, we have treated the quantum degrees of freedom exactly
and the classical variables at the mean-field level. This approach
correctly takes into account effects of disorder on the electronic
sector, due to both the random position of the impurities and the
thermal fluctuation of the individual direction of the spins of the
impurities. Moreover, the approach also takes into account the feedback
of this disorder effects on the effective interaction between the
impurity-spins. 

Assuming that all the adatoms sit on a single graphene sublattice,
a ferromagnetic phase has been found, with a critical temperature
that depends linearly on the adatom concentration $x$, at low $x$.
For an isotropic $s$-$d$ interaction, there is no sign of critical
concentration $x_{c}$ below which ferromagnetism is lost. A critical
$x_{c}$ shows up only when the coupling in the $xy$ plane is made
stronger than the coupling in the $z$ direction. For a fixed $x$,
we have determined the variation of the ferromagnetic critical temperature
with charge carrier density, providing proof of concept for a ferromagnetic
transition tunable by electrical means. Regarding the spectral properties
of the electronic system, we have found that in the ferromagnetic
phase, the spin polarization of the electrons is not balanced around
zero energy. This is due to a pseudogap regime, where one polarization
is strongly suppressed above zero energy, while the other is suppressed
below. Since the suppressed component only contributes through Lifshitz
tails, which are made of localized states, the spin polarization of
charge carriers is expected to be close to 100\% in the pseudogap
region.

Allowing the adatoms to distribute randomly between the two sublattices,
antiferromagnetism sets in, with the impurity spins ordering in opposite
directions in the two sublattices. As in the ferromagnetic case, the
low $x$ critical temperature also depends linearly on $x$. No critical
$x_{c}$ is found below which the antiferromagnetic ordering is lost.
Inside the antiferromagnetic phase, there is always a gap in the spectrum,
despite adatom position and spin-orientation disorders. For adatom
concentrations below $x\sim10\%$, the gap depends linearly on $x$,
but for higher values of $x$, strong deviation due to disorder are
observed.

The results obtained here agree with several experimental observations
where adatom decorated graphene has been shown to have a magnetic
response \citep{xie2011room,miranda13,giesbers2013interface,lanzara2016,Tucek2017}.
Within the framework of the $s$-$d$ interaction, and taking into
account the intrinsic disorder effects, the magnetic phases are found
to be stable down to the lowest concentration of adatoms accessed
here ($x\sim1\%$). In particular, the present theory provides a qualitative
understanding for the results of Ref.~\citep{lanzara2016}, where
a ferromagnetic phase has been found below $\sim30\,\text{K}$ for
graphene decorated with S-atoms.

\begin{acknowledgments}
Partial support from FCT-Portugal through Grant No. UID/CTM/04540/2013
is acknowledged. B.A. received funding from the European Union\textquoteright s
Horizon 2020 research and innovation program under the Grant Agreement
No. 706538.
\end{acknowledgments}

\bibliographystyle{apsrev4-1}
\bibliography{draft_FJSousa}

\end{document}